\documentclass[12pt]{article}
\begin{document}
\setcounter{page}{0}
\thispagestyle{empty}
\setlength{\parindent}{1.0em}
\begin{flushright}

\end{flushright}
\renewcommand{\thefootnote}{\fnsymbol{footnote}}
\begin{center}{\LARGE{{\bf A Not So Little Higgs? }}}
\end{center}
\begin{center}{\large{D G Sutherland}}\\
\end{center}
\begin{center}{{\it Department of Physics and Astronomy}\\{\it
University of Glasgow, Glasgow G12 8QQ, Scotland}}\end{center}
 
\renewcommand{\thefootnote}{\arabic{footnote}}
\setcounter{footnote}{0}

\begin{center}
{\bf Abstract}
\end{center}

{\small Most recent models assuming the Higgs Boson is a
  pseudo-Nambu-Goldstone Boson (pNGb) are motivated by the indication
  from Standard Model fits that its mass is $\leq 200 GeV$ .Starting
  from a modified SM. of Forshaw et. al. with a triplet boson added
  and a heavier Higgs Boson, we consider a pNGb model.This differs in
  several ways from most little Higgs models: apart from using only
  one loop, the cutoff scale is reduced to 5 TeV, and consequently a
  linear sigma model is used to alleviate FCNC effects; no new vector
  bosons are required, but vector-like isosinglet fermions are needed,
  but play no part in determining the mass of the Higgs boson.The
  phenomenology of the isosinglet pNGb that arises from the $SU(3)
  \times SU(3) \rightarrow SU(3)$ model we use is briefly discussed.
  Some potential theoretical and phenomenological problems are
  mentioned briefly.

Tel:0141 330 5160

Fax:0141 330 6436

e-mail:d.sutherland@physics.gla.ac.uk



\newpage

\section{Introduction}

The indication from Standard Model (SM) fits to precision data that
the mass of the Higgs Boson is $\leq 200 GeV $ has motivated many
recent, and often ingenious, models, the Little Higgs Models (LHMs).
For reviews see \cite{rev}. Typically these models assume a global
symmetry group at $\simeq$ 10 TeV which breaks spontaneously to give
Nambu-Goldstone bosons amongst which are the Higgs Bosons. These
acquire mass from radiative corrections, but the models are
constructed so that the one loop quadratic divergences cancel, thereby
ensuring a light enough Higgs Boson.  \par
Experimentally, however, there is only a lower bound on the mass of
the Higgs Boson. Soon after the precision data appeared several
authors \cite{smdiff} considered how the limit on the mass could be
raised by modest alterations of the SM. Amongzt these was a model due
to Forshaw and collaborators\cite{forcol} They showed that by adding a
real triplet scalar boson with a small vacuum expectation value
adequate fits to precision data with a Higgs Boson mass of 500 GeV
(and similar maas for the triplet) coud be obtained.
\par
This suggests the possibility of a model where the Higgs Boson is a
pseudo-Nambu Goldstone boson (pNGb), but the global group is taken at
5 TeV, and, since 0.5 TeV$\simeq$ $\sqrt\alpha\times 5$TeV there may
be no need for extra geuge bosons, or fermions to ensure the
cancellation of divergences. It transpires that it is possible to
eliminate the need for extra gauge bosons, but extra fermions seem
necessary, but are not constrained by contributing to the mass of the
Higgs boson as in many LHMs.
\par                                    

The model is presented in the next section with a particular emphasis
on the need to use a linear, as opposed to the non-linear sigma model
generally used in LHMs. The next section gives the
Coleman-Weinberg\cite{colwei} potential of the model, The
Coleman-Weinberg potential for the isoscaler partner $\eta$ of the
Higgs is given in the next section, and the phenomenology of the
$\eta$ is discussed briefly. In the final section some open problems
which remain to be resolved are discussed, and a conclusion given.

\section{The Model}

Forshaw et al add a real triplet scalar field to the SM. One must then
look for a group whose breaking will produce a triplet $\phi^i$, a
complex doublet $H^a$ and possibly some singlets as commonly arise in
addition in LHMs. Without considering product groups no candidate has
appeared, but the group $ SU(3) \times SU(3)$ which breaks to SU(3)
seems well suited to this purpose, and gives just one singlet $\eta$.

\par

Unlike most LHMs a linear rather than non-linear sigma model is used.
There are three reasons for this. First a comparison with Forshaw
et.al.'s field theoretical analysis would be difficult for the
non-linear case as higher powers of fields suppressed only by powers
of the breaking scale f$\simeq 0.5 TeV$ would appear. Secondly recall
the old paper of Georgi and Kaplan\cite{geokap} who used this same
group with a non-linear sigma model, but felt dissatisfied as
precision tests required f too large, a view strengthened now by f
being $\geq3 TeV$\cite{pespei} Georgi and Kaplan did not consider a
small triplet vev so that one might think that allowing this could
improve the situation, but by using their exponential parametrisattion
one finds that the triplet vev and the 'effective triplet vev'
O($v^2/f^2$) where v is vev of $H^0$ are out of phase by $\pi/2$, so
that the problem is made worse.

\par

 A third reason comes from the constraints of FCNC. Chivukula et
al.\cite{chivev} have argued that these constraints require a cutoff
scale well above the 10 TeV of LHMs. Clearly if one lowers the scale
to 5 TeV this problem becomes more serious. One remedy
suggested\cite{arkanel} is to have the LH as a linear sigma model
which arises as a little Higgs model from a scale an order of
magnitude higher. Such an idea has recently been implemented for the $
SU(3) \times U(1) $ LHM\cite{kapsm}. This again suggests the use of a
linear sigma model, though it has to be stressed that no UV completion
has yet been obtained for the $ SU(3) \times SU(3) $ model.

\par

Extra fermions, singlets under SU(2), will now appear to fill triplets
along with t and b quarks, as well as along with lighter quark
multiplets.The extra singlets can give rise to FCNC problems by mixing
with quarks of the first two generations. This has recently been
analysed by Deshpande et al.\cite{deshpa} who find the strong
constraint $|U_{ds}|\leq1.2.10^{-5}$ from rare K decays in a model
wiith an extra charge -1/3 quark, where $U_{ds}$ denotes the mixing
between d and s induced by the extra quarks. Provided the singlet
quarks are heavy, and the decreasing mixing between light and heavy
quarks seen in the SM can be extended to new quarks, this constraint
may (just) be satisfied.

\par

\section{The Coleman-Weinberg Potential for $\phi$ and H.}

The scalar potential used by Forshaw et al is given, in our notation, by
\begin{equation}
  \mu_1^2|H^2| +\mu_2^2/2|\phi^2| + \lambda_1|H^4| + \lambda_2/4|\phi^4| + \lambda_3/2|H^2| |\phi^2| + L_3
  \label{forpot}
\end{equation}
where 
\begin{equation}
L_3 = \lambda_4\phi^iH^\dagger\sigma^i H
\label{Ldef}
\end{equation}

\par

One can ask how much of this potential can be produced by a
Coleman-Weinberg mechanism. The Coleman-Weinberg potential gives rise
to quadratically divergent coefficients of $\phi^2$ and $H^2$, as well
as logarithmic divergences for $\phi H^2$ and terms quartic in $\phi$
aand H. The $\phi H^2$ term is novel and such a term will not arise in
the Coleman Weinberg effective potential generated using only SM gauge
bosons and fermion loops. This is because the gauge bosons couple to
bilinears in $\phi$, while doublet fermions and right handed singlet
fermions do not couple to the isovector $\phi $. As is shown below the
terms in $ \lambda_1, \lambda_2$, and $\lambda_3$ are also
inadequately described by the Coleman-Weinberg potential so that only
the terms in $ \mu_1$ and $ \mu_2$ can be treated, that is the terms
which are directly related to the Goldstone origin of H and $ \phi.$

\par

The quadratically divergent $\phi^2$ term is given by
\begin{equation}
V(\phi^2) = 3  g_2^2/32\pi^2\Lambda^2
\label{phiequ}
\end{equation}
from gauge bosons. For $\Lambda= 5TeV$ the (positive) mass squared
=0.2$ TeV^2$ for $\phi$.  For H the dominant (negative) mass squared
is expected to come from the top quark loop and is of magnitude
$\simeq 2 TeV^2$. The positive contribution of gauge bosons is small
$\simeq 0.2 TeV^2$, but unlike the case of light H a large positive
contribution comes from H loop itself. For $ m_H =0.5 TeV$ this is
given by $\frac{\lambda_1 \Lambda^2}{8\pi^2}$ where $\lambda_1$
$\simeq 2 $ for $m_H=0.5TeV$. This gives a mass squared of $\simeq
0.63 TeV^2$. Furthermore $\lambda_3\simeq\lambda_1$ typically in the solutions of
Forshaw et al with heavy scalars so that this will give a further
positive contribution of similar magnitude. Also, as can be seen from the next section, a similar positive contribution can be expected from the $\eta$ loop, though this is more uncertain. Taken together with the contribution of the gauge bosons this could have the disastrous effect of making $\mu_1^2$ positive. This problem can be resolved either by noting that each term is only given up to a
constant of O(1) from UV uncertainties or by having $m_H$ somewhat less than 500 GeV when the contributions of the scalar loops are reduced by $(m_H/500GeV)^2$so that a negative value of the corrsct magnitude may be obtained for
$\mu_1^2$. The second approach is favoured by the existence of many  
more solutions for $m_H$ somewhat less than 500 GeV than for $m_H=500GeV$, but, because of the uncertainty associated with the first approach, one can hardly
regard it as a prediction of $m_H$ of the model.

\par

Because of the large $\lambda_3$ there will be a significant positive
contribution to $\mu_2^2$. For $m_H =500 GeV$ $\mu_2^2/2\simeq 0.5 TeV^2$,
much as desired, but the $\lambda_2$ term would give a further
positive contribuion, which is hard to determine from\cite{kapsm}.Thus
there may be a need to invoke the first approach for $\mu_2^2$,
Another possibility is to add some bare term , presumably coming from
some still higher scale, as for $m_\pi$ in QCD, as done in the
SU(3)xU(1) little Higgs model\cite{Schmaltz}.

\par

Overall it appears that fair consistency at least can be achieved with
the Coleman-Weinberg mechanism for the quadratic terms in the model,
although some uncertainty still remains. The situation is quite
different for the quartic terms. As mentioned in the previous
paragraph $\lambda_1$ must be $\simeq 2$ to achieve $m_H\simeq 0.5 TeV$
as desired, but the Coleman-Weinberg value $\simeq
\frac{3log(\Lambda/m_t)}{8\pi^2}$ dominantly from the box diagram from
the top quark,where the fact that the Yukawa coupling of the top quark $\simeq 1$ has been used, but with a substantial reduction in magnitude coming
from gauge and H boson box diagrams, Even neglecting these, one finds
$\lambda\simeq 0.12$, far short of 2. Thus it seems impossible to
accommodate a heavy Higgs boson purely within the scheme of a
radiatively generated Higgs potential, It is clear, however, from the
paper of Coleman-Weinberg that quartic tree interaction is allowed of
a priori undetermined magnitude, although one may be uneasy that it is
an order of magnitude bigger than the radiatively generated one.

\par

A similar problem will arise for the $\lambda_2$ and $\lambda_3$ terms
of Forshaw et al's model.They cannot be much bigger than $\lambda_1$
as obtained by the Coleman-Weinberg mechanism since $g_2^2$
$\leq\lambda_t^2$ where $ \lambda_t $ is the top quark coupling to H
and $ g_2$ is the coupling of gauge bosons to $\phi$. In any case the
necessity of dominant tree contributions is most apparent for
$HH^\dagger HH^\dagger$ interactions. Of course, once one invokes
large tree terms there is no reason why they should not appear in any
term in the potential (beyond the quadratic or there is nothing to
discuss).\footnote{A model has been constructed\cite{suthe}
  introducing vector-like fermions of mass $\simeq5$TeV in an
  adaptation of a model due to Popovic\cite{Popovich}. While this can
  reproduce radiatively $ L_3$ to an isospin conserving accuracy of a
  few per cent, there seems little advantage in this complexity, which
  could be regarded as an attempt to second guess dynamics at the
  cutoff, once one fails to obtain other terms in the potential in
  this way.}
 
\par

\section{Phenomenology of $\eta$}

Recalling that the model has an octet of pNGb's consisting of the complex doublet Higgs boson, an isotriplet and an isosinglet $\eta$, the phenomenology of the $\eta$ has to be examined to ensure that it causes no problems,The potential for $\eta$ has the form (cf.the potential of Forshaw et.al.)
\begin{equation}
\lambda_{2,\eta}/4 |\eta^4| + \lambda_{3,\eta}/2 |\eta^2||H^2| + L_\eta
\label{etapot}
\end{equation}
where
\begin{equation}
L_\eta=\lambda_{4,\eta}\eta H^\dagger H  + \lambda_{5,\eta}\eta^3 + \lambda_{6,\eta}|\eta^2| |\phi^2|
\label{Lthreeequ}
\end{equation}
Here $\lambda_{4,\eta}$ is given by d-type SU(3) coupling as
$\sqrt(2/3) \lambda_4$. There are no terms such as $\mu_1$ and $\mu_2$
$\propto\Lambda^2$ from gauge and fermion loops, but mass will be
induced from $\lambda_{3,\eta}$, $\lambda_{6,\eta}$, and
$\lambda_{2,\eta}$ terms, both $\propto \Lambda^2$ and from the
respective vevs.
 
\par 

Because of the term linear in $\eta$ a vev will be induced for $\eta$
in similar fashion to that for the littlest Higgs model\cite{littlest}
and here for $\phi$. It is expected to be much smaller than $\langle H
\rangle$ as was $\langle \phi^0 \rangle$. Other than slightly
aggravating the already uncontrolled problem of the cosmological
constant, it is not clear what consequences such a vev has.

\par
 
The mass of $\eta$ is given, as above, by loops of H ,$\phi$ and
$\eta$ itself, as well as from vevs, dominantly of H, although these
contribute a small amount relative to the uncertainties from loops.
$\lambda_{3,\eta}$, $\lambda_{6,\eta}$ and $\lambda_{2,\eta}$ are not
fully fixed by symmetry from $\lambda$'s, but it seems likely they
will also be O(1) and thereby induce a mass for $\eta\simeq m_\phi$ or
possibly somewhat smaller as the gauge bosons do not contribute.
\par

Because $\eta$ couples only to Higgs pairs amongst SM particles it
seems to require a detailed analysis , beyond the scope of this paper,
to give a reliable estimate of its production cross section. However,
it seems certain that, involving a Higgs-Higgs collision, and if it
weighs several hundred GeV, one can be confident that it would not
have been detected in present experiments.

\par

Being neutral its future detection is likely to be stongly dependent
on its lifetime as well as its decay modes. If $m_\eta\simeq400GeV$,
say, its main decay mode should be top, antitop pairs with a Yukawa
coupling constant $\gamma$, which comes from evaluating a loop with t
exchange between Higgses from the $\lambda_{4,\eta}$ coupling.
$\lambda_{4,\eta}$ can be estimated as follows via constraining
$\lambda_4$.

\par

$\langle\phi^0\rangle/\langle H \rangle$ is bounded from precision
tests by 0.025. From the equation for the minimum of the potential
\begin{equation}
m_\phi^2 \langle\phi^0\rangle = \langle H^2 \rangle \lambda_4
\label{phivacequ}
\end{equation}
one obtains $\lambda_4\leq0.035TeV$ for $m_\phi = 0.5TeV$. While this
is only a bound , one expects $\lambda_4$ not to be substantially less
than this.

\par

Evaluating the triangle loop, assuming $m_\eta$ sufficiently heavy for
decay to t$\bar t$, gives
\begin{equation}
\gamma = \frac{\sqrt(2/3) \lambda_4 m_t}{4\pi^2 m_H^2 (1 + O(m_t^2/m_H^2)}
\label{gamequ}
\end{equation}
Taking for illustration $m_\eta = 400GeV$ one obtains 
\begin{equation}
\Gamma_\eta = \frac{\gamma^2 k}{4\pi}
\label{widequ}
\end{equation}
where k = 130GeV is the momentum of t in the $\eta$ rest frame. With
$|\gamma|\leq 6.6.10^{-4}$ from Eq(\ref{gamequ}) one obtains
$\Gamma_\eta\leq5000 eV$. While this is much less than the width of a
SM Higgs boson of the same mass, (and by the same token its production
cross section is drastically suppressed compared to that of a Higgs
boson) it is too large to give a displaced vertex. If $150GeV\leq
m_\eta\leq360GeV$ $\eta$ will decay to vector boson pairs, and the
lifetime will increase by $O(1/g_2^4)$ or O(10), but still with no
displacement. If $m_\eta\leq 150GeV$ the decay to $b\bar b$ will be
suppressd by $O(m_t^3/m_b^3)$ compared to the case of $m_\eta$ =
400GeV, so that $\tau_\eta$ could be $0(10^{-13})$s. Such a light
$\eta$ seems, however, unlikely, and even this $\tau_\eta$ is probably
too short to give a detectable displaced vertex. Thus $\eta$ is
similar to H in at least its main decay modes, but its production
cross section is so small that it will occasion no confusion with H.
Indeed it is hard to see how it would be produced at any accelerator
in the foreseeable future.

\par

\section{Conclusion}

An approach to resolving the little hierarchy problem using pNGb's has
been presented which does not need new gauge bosons, but at the
expense of extra scalar bosons, an isovector and an isovector, though
the latter appears very hard to detect. The model gives reasonable
masses for the scalars via the Coleman-Weinberg mechanism, though
$m_\phi$ tends to be rather large. The interactions of the scalars
have to come from tree level, since the interactions generated by the
Coleman-Weinberg mechanism are too weak. It is not clear how serious
this is. In QCD the $\pi\pi$ interaction is not usually obtained from
a Coleman-Weinberg mechanism, though pions are prototypes of pNGb's,
and yet it has a $\sigma$ resonance at $\simeq$500MeV. It should be
noted, however, that the dynamics of the SU(3)xSU(3) employed in this
model cannot be similar to that of chiral SU(3)xSU(3) as the
interaction of Eq(\ref{forpot}) are of non-derivative type.

From a theoretical standpoint there are several issues that remain to
be resolved. It is not clear if a Little Higgs model at $\simeq70TeV$
can be constructed so as to give a linear sigma model with SU(3)xSU(3)
symmetry as used here. From\cite{kapsm} it appears that this may prove
very hard.
 
Forshaw et al require in their renormalization analysis that the
$\lambda$s do not become too large by 1 TeV scale. One might worry
that this scale should be extended to 5 TeV here, which would probably
limit further the scalar masses allowed.\footnote{ A
  calculation\cite{espin} of the potential in the littlest Higgs model
  may cast some doubt, however, on the necessity of this.} A further
issue, possibly related to this, is unitarity\cite{uniprob}. These
authors find, in non-linear realisations of pNGbs, that unitarity is
violated below $\Lambda$ given by $4\pi f$ when there are many pNGb's.
The implications of this observation for this model remain to be
analysed.

Despite these open theoretical issues, it seems worthwhile to present
this model, because of its simplicity, economy of new states, and
difference in outlook to most current approaches. It is to be hoped
that it may stimulate other , and perhaps better, models along similar
lines. Finally, and especially if $\Lambda$ is reduced somewhat,
following the lines of \cite{uniprob} the intriguing possibility that
the LHC could access the UV dynamics might arise.

\end{document}